\documentclass[doublecol]{epl2}
\title{Feasibility of Experimental Realization of Entangled
Bose-Einstein Condensation\\ {\it  Europhys. Lett. {\bf 86} (2009)
60008} }
\author{Yu Shi\footnote{Email: yushi@fudan.edu.cn}}
\shortauthor{Yu Shi} \institute{Department of Physics, Fudan
University,
Shanghai 200433, China \\
Kavli Institute for Theoretical Physics China at the Chinese Academy
of Sciences, Beijing 100080, China}

\abstract{ We examine the practical feasibility of the experimental
realization of the so-called entangled Bose-Einstein condensation
(BEC), occurring in an entangled state of two atoms of different
species. We demonstrate that if the energy gap remains vanishing,
the entangled BEC persists as the ground state of the concerned
model in a wide parameter regime. We establish the experimental
accessibility of the isotropic point of the effective parameters, in
which the entangled BEC is the exact ground state, as well as the
consistency with the generalized Gross-Pitaevskii equations.  The
transition temperature is estimated. Possible experimental
implementations are discussed in detail. }


\begin{document}
\maketitle

Recently, a novel type of Bose-Einstein condensation (BEC),
characterized by an entangled order parameter, herein simply called
entangled BEC, was found to be the ground state of a mixture of two
species of atoms with spins~\cite{shi}. The entangled BEC, which
occurs in an interspecies two-particle entangled state, is
interesting in several aspects. First, it is not a mean field state
of single atoms, unlike the usual multicomponent BEC, whose ground
state is simply a direct product of wave functions of all the atoms,
with each component described by a same wave function. Second, the
interspecies two-particle entangled state bears some similarity to
Cooper pairing of fermions, but it is not a weakly bound state. In
other words, the entangled order parameter is nonlocal. Third, this
entangled condensate may serve as a source of
Einstein-Podolsky-Rosen pairs, the key components in a lot of
quantum information processes. The nonlocal nature of the order
parameter makes it convenient to be used for this purpose, if
experimentally realized. Fourth, as elaborated below, the entangled
BEC is a new kind of fragmented condensation, i.e., with macroscopic
occupation of more than one single-particle state~\cite{frag}, which
has attracted a lot of attention in recent years~\cite{frag1,ho1}.

The entangled BEC was previously established  as the ground state of
the concerned model, when the effective parameters are so chosen
that the total pseudospin of the system is conserved, in consistency
with the common wisdom that fragmentation is due to spin
symmetry~\cite{ho1}. An important question is how this ground state
can survive in a wide parameter regime, where the Hamiltonian breaks
or lowers the symmetry. This is crucial for experimental
realization. There have been some numerical investigations on this
issue demonstrating the persistence of interspecies
entanglement~\cite{shi}. But there has been a lack of analytical
discussions. Moreover, it has not been clear enough how to actually
implement the entangled BEC in experiments.

In this Letter, we investigate several related issues around the
experimental realization of the entangled BEC. First we consider the
concerned model with a deviation from the symmetric parameter point,
where the entangled BEC is the exact ground state. Treating the
deviation as a symmetry breaking perturbation, we find that for a
finite-volume gas, if the energy gap tends to vanish, then the
ground state approaches the unperturbed entangled BEC. This
establishes a significant parameter regime for the entangled BEC.
Afterwards, we examine the consistency between the isotropic point
of the effective parameters and the generalized Gross-Pitaevskii
equations of the  orbital  wave functions, which are in turn related
to those effective parameters. After a brief estimation of the
transition temperature, we discuss in detail how to experimentally
implement the entangled BEC. In view of recent related experimental
progress, we suggest that it is feasible to experimentally realize
the entangled BEC in an optical trap.

{\bf The model.} -- Consider a dilute gas of two species (or
isotopes) of bosonic atoms in a trap~\cite{shi}. Each atom possesses
an internal degree of freedom represented as a pseudospin with
$z$-component basis states $\uparrow$ and $\downarrow$. Then the
many-body Hamiltonian is ${\cal H} = \sum_{i=a,b}
[\sum_{\sigma=\uparrow,\downarrow} \int d^3r
\psi_{i\sigma}^{\dagger} h_{i\sigma}\psi_{i\sigma} + \frac{1}{2}
\int d^3r \psi_{i\sigma_1}^{\dagger} \psi_{i\sigma_2}^{\dagger}
U^{(ii)}_{\sigma_1\sigma_2\sigma_3\sigma_4} \psi_{i\sigma_3}
\psi_{i\sigma_4}]+ \int d^3r \psi_{a\sigma_1}^{\dagger}
\psi_{b\sigma_2}^{\dagger}
U^{(ab)}_{\sigma_1\sigma_2\sigma_3\sigma_4} \psi_{b\sigma_3}
\psi_{a\sigma_4}$,  where the independent variable $\mathbf{r}$ in
the field operator $\psi_{i\sigma} $, the single particle
Hamiltonian $h_{i\sigma}$ and the interaction
$U^{(ij)}_{\sigma_1\sigma_2\sigma_3\sigma_4}$ are all omitted for
brevity. $h_{i\sigma}=-\hbar^2\nabla_i^2/2m_i+U_{i\sigma}$, where
$U_{i\sigma}$ is the trapping potential.   The s-wave effective
interaction
$U^{(ij)}_{\sigma_1\sigma_2\sigma_3\sigma_4}(\mathbf{r}_i-\mathbf{r}_j')
=(2\pi\hbar^2\xi^{(ij)}_{\sigma_1\sigma_2\sigma_3\sigma_4}/\mu_{ij})
\delta(\mathbf{r}_i-\mathbf{r}_j')$, $(i,j=a,b)$, where
$\mu_{ij}=m_i m_j/(m_i+m_j)$ is the effective mass,
$\xi^{(ij)}_{\sigma_1\sigma_2\sigma_3\sigma_4}$ is  scattering
length for an allowed channel in which the initial pseudospins of
$i$ and $j$ are $\sigma_4$ and  $\sigma_3$, respectively, while
their final pseudospins are  $\sigma_1$ and  $\sigma_2$,
respectively. As usual, in ignorance of the depletion, the orbital
degree of freedom of each atom is constrained in the manifold of the
single-particle orbital ground states
$\phi_{i\sigma}(\mathbf{r}_i)$. Hence
$\psi_{a\sigma}=a_{\sigma}\phi_{a\sigma}(\mathbf{r}_a)$ and
$\psi_{b\sigma}=b_{\sigma}\phi_{b\sigma}(\mathbf{r}_b)$, where
$a_{\sigma}$ and $b_{\sigma}$ are annihilation operators. Under the
conservation of $z$ component of the total pseudospin in each
scattering, the many-body Hamiltonian is simplified to
\begin{eqnarray} {\cal H} =   \sum_{i,\sigma} f_{i\sigma} N_{i\sigma}+
\frac{1}{2}\sum_{i,\sigma\sigma'}K^{(ii)}_{\sigma\sigma'}
N_{i\sigma}N_{i\sigma'}\nonumber  \\
+\sum_{\sigma\sigma'}K^{(ab)}_{\sigma\sigma'}N_{a\sigma}
N_{b\sigma'}  + \frac{K_e}{2} (a^{\dagger}_{\uparrow}a_{\downarrow}
b^{\dagger}_{\downarrow}b_{\uparrow} +
a^{\dagger}_{\downarrow}a_{\uparrow}
b^{\dagger}_{\uparrow}b_{\downarrow}), \label{effective}
\end{eqnarray}
where $N_{i\sigma}$ is the number of atoms of species
$i$ with pseudospin $\sigma$, $N_i = N_{i\uparrow} +N_{i\downarrow}$
is conserved, $f_{i\sigma} \equiv \epsilon_{i\sigma}-
K^{(ii)}_{\sigma\sigma}/2,$
\begin{equation} \epsilon_{i\sigma}=\int \phi_{i\sigma}^*
(-\hbar^2\nabla_i^2/2m_i+U_{i\sigma})\phi_{i\sigma} d^3r \label{e}
\end{equation} is the single particle energy of an atom of species
$i$ and pseudospin $\sigma$.   Note that there is no a priori
requirement on the relation between $\phi_{i\uparrow}$ and
$\phi_{i\downarrow}$. $K_e$ and $K^{(ij)}_{\sigma\sigma'}$ are
effective parameters proportional to the corresponding scattering
lengths, and are defined in the following way. First, for the
scattering in which an $i$-atom flips from $\sigma_4$ to $\sigma_1$
while a $j$-atom flips from $\sigma_3$ to $\sigma_2$, define
\begin{equation}
K^{(ij)}_{\sigma_1\sigma_2\sigma_3\sigma_4}\equiv
\frac{2\pi\hbar^2\xi^{(ij)}_{\sigma_1\sigma_2\sigma_3\sigma_4}}
{\mu_{ij}}\int \phi_{i\sigma_1}^*\phi_{j\sigma_2}^*
\phi_{j\sigma_3}\phi_{i\sigma_4} d^3r. \label{ks}
\end{equation} Then in accordance with the convention of
Leggett~\cite{legrev}, we use the shorthands
$K^{(ii)}_{\sigma\sigma}\equiv K^{(ii)}_{\sigma\sigma\sigma\sigma}$,
and $K^{(ii)}_{\sigma\bar{\sigma}}\equiv
2K^{(ii)}_{\sigma\bar{\sigma}\bar{\sigma}\sigma} =
2K^{(ii)}_{\sigma\bar{\sigma}\sigma\bar{\sigma}}$ for $\sigma\neq
\bar{\sigma}$ for intraspecies scattering, while
$K^{(ab)}_{\sigma\sigma'}\equiv
K^{(ab)}_{\sigma\sigma'\sigma'\sigma}$ for interspecies scattering.
For interspecies pseudospin exchange scattering, we denote
$K_{e}\equiv 2K^{(ab)}_{\uparrow\downarrow\uparrow\downarrow}
=2K^{(ab)}_{\downarrow\uparrow\downarrow\uparrow}$. The last term of
the Hamiltonian (\ref{effective}) represents pseudospin-exchange
scattering between two atoms of different species. Without loss of
generality, suppose $N_a \geq N_b$.

{\bf Ground State and fragmentation.} -- Using the total pseudospins
of the two species, $$\mathbf{S}_a \equiv \sum_{\sigma,\sigma'}
a_{\sigma}^{\dagger} \mathbf{s}_{\sigma\sigma'}a_{\sigma'}, \mbox{ }
\mathbf{S}_b \equiv \sum_{\sigma,\sigma'} b_{\sigma}^{\dagger}
\mathbf{s}_{\sigma\sigma'}b_{\sigma'},$$  and subtracting  a
constant, the Hamiltonian can be rewritten as
\begin{equation}
{\cal H} = {\cal H}_0 + {\cal H}_1, \label{hamiltonian}
\end{equation}
with
$${\cal H}_0 = K_e \mathbf{S}_{a}\cdot\mathbf{S}_{b},$$
$${\cal H}_1 = (J_z-K_e) S_{az} S_{bz} + B_a S_{az} + B_b S_{bz} +C_a
S_{az}^2 +C_b S_{bz}^2,
$$
where
$$\begin{array}{rcl}
J_z&=&K_{\uparrow\uparrow}^{(ab)}+K_{\downarrow\downarrow}^{(ab)}
-K_{\uparrow\downarrow}^{(ab)}-K_{\downarrow\uparrow}^{(ab)},\\
B_a&=&f_{a\uparrow}-f_{a\downarrow}+\frac{N_a}{2}(K_{\uparrow\uparrow}^{(aa)}
-K_{\downarrow\downarrow}^{(aa)})\\
&&+\frac{N_b}{2}(K_{\uparrow\uparrow}^{(ab)}
+K_{\uparrow\downarrow}^{(ab)} -K_{\downarrow\uparrow}^{(ab)}
-K_{\downarrow\downarrow}^{(ab)}),
\\B_b&=&f_{b\uparrow}-f_{b\downarrow}+\frac{N_b}{2}(K_{\uparrow\uparrow}^{(bb)}
-K_{\downarrow\downarrow}^{(bb)})+
\\&& \frac{N_a}{2}(K_{\uparrow\uparrow}^{(ab)}
+K_{\downarrow\uparrow}^{(ab)}
-K_{\uparrow\downarrow}^{(ab)}-K_{\downarrow\downarrow}^{(ab)}),\\
C_a &=& \frac{1}{2}(K_{\uparrow\uparrow}^{(aa)}
+K_{\downarrow\downarrow}^{(aa)}-K_{\uparrow\downarrow}^{(aa)}
-K_{\downarrow\uparrow}^{(aa)}),
\\ C_b &=&
\frac{1}{2}(K_{\uparrow\uparrow}^{(bb)}
+K_{\downarrow\downarrow}^{(bb)}-K_{\uparrow\downarrow}^{(bb)}
-K_{\downarrow\uparrow}^{(bb)}). \end{array}$$

${\cal H}_0$ is symmetric under $SU(2S+1)$, where $S$ is the total
spin quantum number. The eigenstates of ${\cal H}_0$ are
$|S,S_z\rangle$'s, where $S=S_a-S_b, \cdots, S_a+S_b$, $S_z = -S,
\cdots, S$ is the $z$-component of the total pseudospin,
$S_a=N_a/2$, $S_b=N_b/2$. Obviously, except the ground state in the
case of $N_a=N_b$, the eigenstates are degenerate with respect to
$S_z$.

The ground state of ${\cal H}_0$, in the sector of $S_z$, is
\begin{equation} |G_{S_z}\rangle = {\cal
A}(a_{\uparrow}^{\dagger})^{n_{\uparrow}}
(a_{\downarrow}^{\dagger})^{n_{\downarrow}}
(a_{\uparrow}^{\dagger}b_{\downarrow}^{\dagger}-
a_{\downarrow}^{\dagger}b_{\uparrow}^{\dagger})^{N_b}|0\rangle,
\label{gs1}
\end{equation}
for which $S=S_a-S_b=(N_a-N_b)/2$, ${\cal A}$ is the normalization
constant, $n_{\uparrow}= N_a/2-N_b/2+S_z$, $n_{\downarrow}=
N_a/2-N_b/2-S_z$, $S_z=-(N_a-N_b)/2, -(N_a-N_b)/2+1, \cdots,
(N_a-N_b)/2$ is a conserved quantity.

The ground state $|G_{S_z}\rangle$ is a special kind of fragmented
condensate. Because $a$ and $b$ are two different species of atoms,
one should define a one-particle reduced density matrix for each
species respectively. For species $i$, it is $$\rho_i
(\mathbf{r},\mathbf{r}') \equiv \langle
G|\hat{\psi}_i^{\dagger}(\mathbf{r}')
\hat{\psi}_i(\mathbf{r})|G\rangle,$$ where
$$\hat{\psi}_a(\mathbf{r}) =
\hat{a}_{\uparrow}\phi_{a\uparrow}(\mathbf{r})|\uparrow\rangle+
\hat{a}_{\downarrow}\phi_{a\downarrow}(\mathbf{r})|\downarrow\rangle,$$
$$\hat{\psi}_b(\mathbf{r}) =
\hat{b}_{\uparrow}\phi_{b\uparrow}(\mathbf{r})|\uparrow\rangle+
\hat{b}_{\downarrow}\phi_{b\downarrow}(\mathbf{r})|\downarrow\rangle.$$
It is evaluated that
$$\begin{array}{rcl}
\rho_{a} (\mathbf{r},\mathbf{r}')  & = &
(N_a/2+S_z)\phi^*_{a\uparrow}(\mathbf{r}')
\phi_{a\uparrow}(\mathbf{r}) \\
&& +(N_a/2-S_z)\phi^*_{a\downarrow}(\mathbf{r}')
\phi_{a\downarrow}(\mathbf{r}), \\\rho_{b} (\mathbf{r},\mathbf{r}')
& = &  N_b/2\phi^*_{b\uparrow}(\mathbf{r}')
\phi_{b\uparrow}(\mathbf{r}) +N_b/2\phi^*_{b\downarrow}(\mathbf{r}')
\phi_{b\downarrow}(\mathbf{r}).   \end{array} $$
Therefore, all
atoms of species $a$ form a fragmented condensate while all atoms of
species $b$ form another one. There is one-particle off-diagonal
long-range order (ODLRO) within each species.

Note that in our definition of the field operators and reduced
density matrices, we have included summation over the two spin
states, i.e. we have used a spinor field operator, in considering
the coherence between the two spin states. The reduced density
matrix of each species characterize the position order for all the
atoms of this species.

The interspecies entanglement leads to nonvanishing two-particle
reduced density matrix of two particles of different species,
$$\langle G_{S_z}|\hat{\psi}_b^{\dagger}(\mathbf{r}_b')
\hat{\psi}_a^{\dagger}(\mathbf{r}_a') \hat{\psi}_a(\mathbf{r}_a)
\hat{\psi}_b(\mathbf{r}_b)|G_{S_z}\rangle,$$ which has more than
four eigenvalues of order $N^2$. Hence there is ODLRO in the
interspecies two-particle density matrix. Moreover, there is phase
coherence in the interspecies pairs.

The state (\ref{gs1}) is the so-called  BEC with an entangled order
parameter, simply called entangled BEC. Note that there is ODLRO in
the single particle density matrix of each species of atoms.

{\bf Persistence of entangled BEC.} -- Away from the symmetric point
in the parameter space,  the symmetry breaking perturbation ${\cal
H}_1$ becomes nonvanishing.  ${\cal H}$ still conserves $S_z$,
though $S$ is no longer conserved. Let us focus on the sector of
$S_z=0$, in which any eigenstate of ${\cal H}$ can be expanded as
\begin{equation}
|\Psi_n\rangle = \sum_S \psi_n(S) |S, 0\rangle, \label{gs}
\end{equation}
where the summation is over $S$,  running  from $S_{min} \equiv
|S_a-S_b|$ to $S_a+S_b$.

${\cal H}|\Psi_n\rangle = E_n|\Psi_n\rangle$ leads to
\begin{widetext}
$$(B_a-B_b)\frac{N}{4}[\psi_n(S-1)+ \psi_n(S+1)] +
(C_a+C_b-J_z+K_e)\frac{N^2}{16}[\psi_n(S-2)+2 \psi_n(S)+\psi_n(S+2)]
=  [E_n-E^{(0)}(S)]\psi_n(S),$$
\end{widetext}
where $N \equiv
(N_a+N_b)/2$, $E^{(0)}(S)=K_e S(S+1)$. In deriving this equation, we
have made use of the following consideration. In the sum in
(\ref{gs}), significant contributions come from $S_{min} \leq S \ll
N$ in order to minimize the energy, of which $E^{(0)}(S)$ is the
dominant part. In this range of $S$, it can be obtained that
$\langle S_a S_b S 0|S_a^z|S_a S_b S 0\rangle \approx
\delta_{S,S'-1} N/4+ \delta_{S,S'+1}N/4,$ $\langle S_a S_b S
0|S_b^z|S_a S_b S 0\rangle \approx
-\delta_{S,S'-1}N/4-\delta_{S,S'+1} N/4,$ and
$\langle S_a S_b S 0|(S_i^z)^2|S_a S_b S 0\rangle  \approx
\delta_{S',S+2} N^2/16+\delta_{S',S}N^2/8 +\delta_{S',S-2} N^2/16.$

In the continuum limit,
$\psi_n(S+1)+\psi_n(S-1)-2\psi_n(S)=[\psi_n(S+2)+\psi_n(S-2)-2\psi_n(S)]/4
= d^2\psi_n(S)/dS^2.$ Therefore one obtains
\begin{equation} -\frac{1}{2}\frac{d^2 \psi_n(S)}{dS^2} +
\frac{1}{2}\omega^2 S^2 \psi_n(S) = \nu_n \psi_n (S), \label{u}
\end{equation} where $\omega^2 \equiv 2K_e/d$,
$\nu_n \equiv (E_n+g)/d$, with
$d=(B_b-B_a)N/2-(C_a+C_b+K_e-J_z)N^2/2,$
$g=(B_b-B_a)N/2-(C_a+C_b+K_e-J_z)N^2/4$. If $\omega^2 >0$,
Eq.~(\ref{u}) becomes the Schr\"{o}dinger equation for a simple
harmonic oscillator, whose $n$-th eigenvalue is $\nu_n =
(n+\frac{1}{2})\omega$. Hence the energy level of the original
Hamiltonian (\ref{hamiltonian}) is given by $$E_n= (n+1/2)\sqrt{2K_e
d}-g.$$ The interaction gives rise to a term squared in the
summation variable. This is the reason why the superposition
coefficients behave like eigenfunctions of harmonic oscillators.

Therefore the energy gap of the Hamiltonian (\ref{hamiltonian}) is
\begin{equation} \Delta = \sqrt{2K_e d}.\label{gap}\end{equation}
Thus
$$\omega=2K_e/\Delta.$$

The ground state of ${\cal H}$, in $S_z=0$ sector, is thus
$$|\Psi_0\rangle = \sum_S \psi_0(S)|S,0\rangle, $$ where
$$\psi_0(S)={\cal A} e^{-\frac{1}{2}\omega S^2}= {\cal
A}e^{-\frac{K_e}{\Delta} S^2},$$  where the normalization constant
${\cal A} =\sqrt{\sqrt{\omega}/erfc (\sqrt{\omega}S_{min})}$, with
$erfc(x)\equiv\int_x^{\infty} e^{-x^2} dx$. As $\psi_0(S)$ is
Gaussian, $S_{min}$ term dominates. The smaller the energy gap
$\Delta$, the more dominant $S_{min}$ term. Especially, in case
$N_a=N_b$,  $S_{min}=0$, thus
$$\psi_0(S)=(\frac{4\omega}{\pi})^{1/4} e^{-\frac{1}{2}\omega
S^2}=(\frac{8K_e}{\pi\Delta})^{1/4} e^{-\frac{K_e}{\Delta}S^2}.$$
Hence
$$\psi_0(S) \rightarrow \delta(S), \mbox{ } {\rm as} \mbox{ } \Delta
\rightarrow 0.$$

Therefore, in case $N_a=N_b$, when the energy gap $\Delta$ is
vanishing  while $K_e \propto 1/\Omega$ remains finite, the ground
state of the asymmetric Hamiltonian approaches the symmetric state
$|S=0,S_z=0\rangle$. One can say that the symmetry and the entangled
BEC is protected by the vanishing energy gap.

More generally, no matter whether $N_a=N_b$,  as far as $ \Delta \ll
K_e$, the ground state of the system can be well approximated as
$|S_{min},0\rangle$. Therefore, the entangled BEC persists as the
ground state of ${\cal H}$, despite that the symmetry is broken in
the Hamiltonian.

A key factor leading to this result is that $\Omega$ remains finite.
If, on the contrary, one takes thermodynamic limit
$N\rightarrow\infty$, $\Omega\rightarrow\infty$ while $N/\Omega$
remains constant, then $K_e \rightarrow 0$, consequently $\psi_0(S)$
becomes independent of $S$, even though one takes $\Delta\rightarrow
0$ after taking $\Omega\rightarrow\infty$. Consequently the ground
state becomes an equal superposition of $|S,0\rangle$ of all
possible values of $S$, hence breaks the symmetry. When the symmetry
breaking perturbation is infinitesimal, the situation becomes
spontaneous symmetry breaking (SSB), and is related to Lieb-Mattis
infinite-range model~\cite{lieb}, with each species of atoms in our
model corresponding to a sublattice in the latter.

The present case of persistence of symmetry, and thus fragmentation
and entanglement, in a finite-volume condensate, could be viewed as
a converse case of SSB of the ground state of an infinite system.
Both are due to the near degeneracy of the ground state and the low
lying excited states~\cite{anderson}. For SSB, the limit $\Omega
\rightarrow \infty$ should be taken before $\Delta \rightarrow 0$.
Likewise, for symmetry persistence discussed here, $\Omega$ should
be kept finite before taking $\Delta \rightarrow 0$.

{\bf Consistency of the isotropic parameter point with the
generalized Gross-Pitaevskii equations.} -- The effective parameters
$K$'s are dependent on the orbital wave functions
$\phi_{i\sigma}$'s, as indicated in Eq.~(\ref{ks}). These wave
functions satisfy the four generalized Gross-Pitaevskii equations,
which are in turn derived from the unique ground state $|G_0\rangle$
at the isotropic parameter point~\cite{shi}. The equation for
$a$-atoms with spin $\sigma$ is $
\{-\frac{\hbar^2}{2m_a}\nabla^2+U_{a\sigma}(\mathbf{r})
+[2(N-1)/3]g^{(aa)}_{\sigma\sigma}
|\phi_{a\sigma}(\mathbf{r})|^2+[(N-1)/3]g^{(aa)}_{\sigma
\bar{\sigma}}
|\phi_{a\bar{\sigma}}(\mathbf{r})|^2+[(N-1)/3]g^{(ab)}_{\sigma\sigma}
|\phi_{b\sigma}(\mathbf{r})|^2
+[(2N+1)/3]g^{(ab)}_{\sigma\bar{\sigma}}
|\phi_{b\bar{\sigma}}(\mathbf{r})|^2 \} \phi_{a\sigma}(\mathbf{r})
-[(N+2)/6]g_e \phi^*_{b\bar{\sigma}}(\mathbf{r})
\phi_{b\sigma}(\mathbf{r})\phi_{a\bar{\sigma}}(\mathbf{r}) =
\mu_{a\sigma}\phi_{a\sigma}(\mathbf{r}),$ where $\bar{\sigma}\neq
\sigma$, $\mu_{a\sigma}$ is the corresponding chemical
potential~\footnote{In Ref.~\cite{shi}, as a Lagrange multiplier for
the normalization of $\phi_{a\sigma}$, $\mu_{a\sigma}$ represents
$N/2$ multiplied by the chemical potential}, each $g$ parameter is
the part preceding the integral in the definition of the
corresponding $K$ parameter in Eq.~(\ref{ks}), i.e.
$g^{(ij)}_{\sigma\sigma'}=2\pi\hbar^2\xi^{(ij)}_{\sigma\sigma'}/\mu_{ij}$,
$g_{e}=4\pi\hbar^2\xi^{(ab)}_e/\mu_{ab}$, where
$\xi^{(ab)}_e=\xi^{(ab)}_{\uparrow\downarrow\uparrow\downarrow
}=\xi^{(ab)}_{\downarrow\uparrow\uparrow\downarrow }$. The equation
for $\phi_{b\sigma}(\mathbf{r})$ is in a similar form.

Therefore, in order to attain the isotropic parameter point, both
the orbital wave functions and the scattering lengths need to be
constrained in order that the effective parameters satisfy the
requirements that $B_i=C_i=0$ and $J_z=K_e$.  Because the
generalized Gross-Pitaevskii equations, which govern the orbital
wave functions, are in turn derived at the isotropic parameter
point, a problem arises whether these requirements are consistent
with the generalized Gross-Pitaevskii equations.

Here we show that such consistency is indeed guaranteed under the
following conditions. (i)
$U_{i\uparrow}(\mathbf{r})=U_{i\downarrow}(\mathbf{r})$. (ii) The
intraspecies scattering lengths
$\xi^{(ii)}_{\sigma_1\sigma_2\sigma_3\sigma_4}$'s, for physically
allowed channels, are all equal for each species $i$, denoted as
$\xi_i$. Note that this means
$\xi^{(ii)}_{\sigma\bar{\sigma}}=2\xi^{(ii)}_{\sigma\sigma}=2\xi_i$
in shorthands. (iii) The interspecies scattering lengths satisfy the
relations
$\xi^{(ab)}_{\uparrow\uparrow}=\xi^{(ab)}_{\downarrow\downarrow}$,
denoted as $\xi^{(ab)}_s$, and
$\xi^{(ab)}_{\uparrow\downarrow}=\xi^{(ab)}_{\downarrow\uparrow}$,
denoted as $\xi^{(ab)}_d$, where the subscripts ``s'' and ``d''
represent ``same'' and ``different'', respectively.

First, by considering that the nonlinear terms of interaction vanish
at the boundary of the condensate, it can be seen that
$U_{i\uparrow}(\mathbf{r})=U_{i\downarrow}(\mathbf{r})$ implies
$\mu_{i\uparrow}=\mu_{i\downarrow}$. Then, under the above three
conditions, in considering that as a wave function of bosons,
$\phi_{i\sigma}$ is real, one obtains, from the difference of the
generalized Gross-Pitaevskii equations for $\phi_{a\uparrow}$ and
$\phi_{a\downarrow}$,
$$(\frac{N-1}{3}g_s-\frac{2N+1}{3}g_d)(\frac{\phi_{b\uparrow}}
{\phi_{b\downarrow}}-\frac{\phi_{b\downarrow}}
{\phi_{b\uparrow}})=\frac{N+2}{6}g_e(\frac{\phi_{a\uparrow}}
{\phi_{a\downarrow}}-\frac{\phi_{a\downarrow}}{\phi_{a\uparrow}}),$$
where $g_{s/d}=2\pi\hbar^2\xi^{(ab)}_{s/d}/\mu_{ab}$.  Similarly,
$$(\frac{N-1}{3}g_s-\frac{2N+1}{3}g_d)(\frac{\phi_{a\uparrow}}
{\phi_{a\downarrow}}-\frac{\phi_{a\downarrow}}
{\phi_{a\uparrow}})=\frac{N+2}{6}g_e(\frac{\phi_{b\uparrow}}
{\phi_{b\downarrow}}-\frac{\phi_{b\downarrow}}{\phi_{b\uparrow}}).$$
Consequently, we obtain $\phi_{a\uparrow}=\phi_{a\downarrow}$ and
$\phi_{b\uparrow}=\phi_{b\downarrow}$, unless $(N-1)g_s=(2N+1)g_d$.
Subsequently, $\epsilon_{i\uparrow} = \epsilon_{i\downarrow}$
according to Eq.~(\ref{e}). Consequently, $B_i=C_i=0$ if the
scattering lengths satisfy the latter two conditions above.
Moreover, $J_z=K_e$ if $\xi^{(ab)}_e = \xi^{(ab)}_s-\xi^{(ab)}_d$.

{\bf Transition Temperature.} -- We briefly discuss the issue of
transition temperature. A crude estimation can be made by following
the idea of Ashhab and Leggett on $SU(2)$ symmetric model of one
species of spin-$\frac{1}{2}$ atoms~\cite{ashhab}. One may consider
four independent interpenetrating gases of $N_{i\sigma}$ atoms  of
species $i$ with spin $\sigma$, ($i = a, b$, $\sigma = \uparrow,
\downarrow$). The transition temperature for each gas is
$T_{i\sigma} \approx 3.31\hbar^2 m (N_{i\sigma}/V)^{2/3}$. The error
due to ignoring spin exchanges is of the order of $\sqrt{N}$, and
can be neglected in this estimation. Hence the transition
temperature of the entangled BEC is roughly the minimum among
$T_{i\sigma}$'s. Below this temperature, the noncondensed atoms of
each species with each spin is $\approx (mT/3.31\hbar^2)^{3/2}$

{\bf Experimental implementations.} -- We now consider how to
implement this model by using the trapped alkali atoms. In order to
constrain each atom in the Hilbert space of only two spin states,
scattering to other spin states must be suppressed.

We propose to represent the two pseudospin states as the hyperfine
states $|F=2,m_F=2\rangle$ and $|F=1,m_F=1\rangle$ of an alkali atom
with nuclear spin $I=3/2$ ($^7$Li, $^{23}$Na, $^{39}$K, $^{41}$K and
$^{87}$Rb).  Similarly, one can also use $|F=2,m_F=-2\rangle$ and
$|F=1,m_F=-1\rangle$ when the magnetic field is small so that their
energy difference is large. Because of the large energy splitting
between the two hyperfine states~\cite{pethick},  the number of
atoms with $F=2$ is forbidden to increase in each scattering.
Furthermore, in an energetically allowed scattering channel,
conservations of the total $F$ and $m_F$ of the two atoms guarantee
that the hyperfine states of the two scattered atoms are either
unchanged or exchanged, just as in our model. Besides, because of
its various advantages, e.g. large interspecies scattering
lengths~\cite{burke}, a mixture of isotopes $^{85}$Rb and $^{87}$Rb
can also be used. For $^{85}$Rb, $I=5/2$, consequently
$|F=3,m_F=3\rangle$ and $|F=2,m_F=2\rangle$ play similar roles as
$|F=2,m_F=2\rangle$ and $|F=1,m_F=1\rangle$ of atoms with $I=3/2$,
respectively. Thus these two sets of hyperfine states can represent
the two pseudospin states of the two species in our model.
Similarly, one can use $|F=3,m_F=-3\rangle$ and $|F=2,m_F=-2\rangle$
of $^{85}$Rb while $|F=2,m_F=-2\rangle$ and $|F=1,m_F=-1\rangle$ of
$^{87}$Rb. The number of $^{85}$Rb cannot be too large in order to
avoid collapse, as its intraspecies scattering length is negative.

Magnetic trapping implementation is not favored for the following
reasons. First of all, only one of the two hyperfine states
representing the two pseudospin states is low-field seeker and can
be trapped. Besides, one of the above conditions for the consistency
of isotropic parameter point, namely
$U_{i\uparrow}=U_{i\downarrow}$,  cannot be satisfied, because  the
magnetic trapping potential is proportional to $g_F
m_F$~\cite{pethick}, hence has different values for the two
hyperfine states. Moreover,  it is difficult to tune the scattering
lengths in a magnetic trap.

In contrast, the implementation can be made in an optical trap,
where the trapping potential is based on the single-atom energy
shift due to red detuning in coupling with the laser, and is
independent of atomic spins~\cite{pethick}, thus the two hyperfine
states representing the pseudospin states can both be trapped.
Moreover, $U_{i\uparrow}=U_{i\downarrow}$ is satisfied. Furthermore,
the scattering lengths can be tuned by using Feshbach resonances.
Therefore, the consistency of the isotropic parameter point can be
achieved. Even though the effective parameters deviate from the
exact isotropic point, the energy gap can still be tuned to vanish,
so that the ground state still approaches the entangled BEC.

Let us summarize some relevant experiences obtained from previous
experiments on BEC mixtures.  In general, for BEC in optical traps,
atoms are first prepared in a magnetic trap.  Optical pumping and
adiabatic passage or radiofrequency sweep are often used in
preparing atoms in certain spin states by transferring atoms from
other spin states. These methods can be adopted in preparing our
system. In the early experiments in magnetic traps, on mixtures of
$|F=2,m_F=2\rangle$ and $|F=1,m_F=-1\rangle$ of
$^{87}$Rb~\cite{myatt} and of $|F=2,m_F=1\rangle$ and
$|F=1,m_F=-1\rangle$ of $^{87}$Rb~\cite{hall}, spin exchange
scattering led to atom loss from the magnetic trap.   BEC with
multiple components of different spin states were studied in the
optically trapped $^{23}$Na with  $F=1$~\cite{stenger} and $^{87}$Rb
with $F=1$~\cite{chang}, where spin exchange scattering indeed did
not produce atoms with $F=2$. Spin exchange scattering were also
observed in $^{87}$Rb with $F=2$ in optical traps, where in some
cases, e.g. the two initial hyperfine states are $|F=2,m_F=2\rangle$
and $|F=2,m_F=-2\rangle$, or  both $|F=2,m_F=0\rangle$, scattering
between $F=2$ atoms lead to $F=1$ atoms~\cite{schmaljohann}.  But
such $F=2$ to $F=1$ scattering is prohibited by the conservation of
total $m_F$ if the initial hyperfine states are $|F=2,m_F=2\rangle$
and $|F=2,m_F=1\rangle$, or both $|F=2,m_F=2\rangle$, as in our
proposed implementations.

Two experimental systems of BEC mixtures are very close to, and
might be extended to, the realization of entangled BEC.  One is a
mixture of $^{41}$K and $^{87}$Rb, which was regarded as the most
favorable candidate for realizing BEC mixture~\cite{italy1}. Their
BEC mixture has been realized in a magnetic trap, with both species
in $|F=2,m_F=2\rangle$~\cite{italy2}, as well as in an optical trap,
with both species in $|F=1,m_F=1\rangle$~\cite{italy3}. In preparing
the latter experiment, atoms are transferred to $|F=1,m_F=1\rangle$
from $|F=2,m_F=2\rangle$ by applying a microwave and  a
radiofrequency sweep. This procedure could also be used for our
purpose, with only half of the atoms transferred. Interspecies
Feshbach resonances, as needed also for our system, were used in
this experiment. The other experimental system close to ours is that
of $^{85}$Rb in $|F=2,m_F=-2\rangle$ and $^{87}$Rb in
$|F=1,m_F=-1\rangle$, in which Feshbach resonances were used in
creating heteronuclear molecules~\cite{wieman1} and in tuning the
two-species BEC~\cite{wieman2}. To realize entangled BEC, we need
also $^{85}$Rb in $|F=3,m_F=-3\rangle$ and  $^{87}$Rb in
$|F=2,m_F=-2\rangle$, which can be transferred from the other  spin
state of each isotope by optical pumping or radiofrequency sweep.

As an alternative approach, the two pseudospin states could also be
implemented by using atoms with negligible hyperfine
coupling~\cite{hulet}.

To prepare an entangled BEC, one may first prepare four condensates
of equal numbers of atoms, of the two species with the two spin
states, as prescribed above. Then conservations of the total spin
and its $z$-component of any two scattered atoms dictate that these
atoms remain in the manifold of these spin states. The interaction
between these four condensates builds up coherence among them, and
realizes entangled BEC of the total system.

{\bf Summary.} -- To summarize,  we have examined the feasibility of
experimental realization of the entangled BEC, i.e. BEC occurring in
an interspecies entangled two-particle state, as the ground state of
a model of two species of atoms with spins, proposed in
Ref.~\cite{shi}. The entangled BEC is a novel type of fragmented
BEC. We have analytically shown that in a wide parameter regime, the
entangled BEC persists as the ground state of the concerned model,
as far as the energy gap tends to vanish. This makes the entangled
BEC more accessible in experiments. Subsequently, we established the
consistency between the isotropic point of the effective parameters
and the generalized Gross-Pitaevskii equations, which govern the
orbital wave functions, on which the effective parameters depend. A
brief estimation of the transition temperature was made. Finally, we
discussed how to experimentally realize this model with the
entangled BEC as its ground state. We found that it is suitable to
use an optical trap. Using two species of atoms with $I=3/2$, the
two pseudospin states of the two species can be realized both in
terms of $|F=2,m_F=2\rangle$ and $|F=1,m_F=1\rangle$, or both in
terms of $|F=2,m_F=-2\rangle$ and $|F=1,m_F=-1\rangle$. In view of
previous experiments, a most favorable candidate is the mixture of
$^{41}$K and $^{87}$Rb. Another candidate is the mixture of
$^{87}$Rb and $^{85}$Rb.  For $^{85}$Rb, $I=5/2$, thus the two
pseudospin states are represented by $|F=3,m_F=3\rangle$ and
$|F=2,m_F=2\rangle$ when $|F=2,m_F=2\rangle$ and $|F=1,m_F=1\rangle$
are used for $^{87}$Rb, or by $|F=3,m_F=-3\rangle$ and
$|F=2,m_F=-2\rangle$ when $|F=2,m_F=-2\rangle$ and
$|F=1,m_F=-1\rangle$ are used for $^{87}$Rb.

\bigskip

I thank Rukuan Wu for discussions. This work is supported by
National Science Foundation of China (Grant No. 10674030) and
Shuguang Project (Grant No. 07S402).

\end{document}